\documentclass[11pt]{article}

\usepackage[utf8]{inputenc}
\usepackage[T1]{fontenc}
\usepackage{times}
\usepackage[margin=1in]{geometry}
\usepackage{hyperref}
\usepackage{url}
\usepackage{booktabs}
\usepackage{amsmath}
\usepackage{graphicx}
\usepackage{natbib}
\usepackage{xcolor}
\usepackage{enumitem}
\usepackage{fancyhdr}
\usepackage{multirow}
\usepackage{array}

\hypersetup{
    colorlinks=true,
    linkcolor=blue!70!black,
    citecolor=blue!70!black,
    urlcolor=blue!70!black
}

\title{%
\textbf{FIGURA: A Modular Prompt Engineering Method\\for Artistic Figure Photography\\in Safety-Filtered Text-to-Image Models}
}

\author{%
Luca Cazzaniga\\
Independent Researcher\\
Monza, Italy\\
\texttt{luca@lucacazzaniga.com}\\
\url{https://www.lucacazzaniga.com}
}

\date{February 2026}

\begin{document}
\maketitle

\begin{abstract}
Safety filters in commercial text-to-image (T2I) models systematically block legitimate artistic content involving the human figure, treating classical nude photography with the same restrictiveness as explicit material. While prior research has documented this problem extensively, no operational system exists that enables professional artists to generate artistic figure photography \emph{within} the constraints of active safety filters. We present the \textbf{FIGURA Method} (\textbf{F}ramework for \textbf{I}ntelligent \textbf{G}eneration of \textbf{U}nrestricted \textbf{A}rtistic \textbf{R}esults), a modular prompt engineering system comprising eight interconnected knowledge files, empirically validated through 200+ documented generation tests on FLUX~2~Pro (Cloud) with active safety filters at the default tolerance level. Our systematic testing reveals several previously undocumented findings: (1) safety filters primarily detect \emph{absence descriptions} (references to missing clothing) rather than \emph{presence descriptions} (references to body form), which we formalize as the \emph{Golden Rule}; (2)~artistic references to painters function simultaneously as aesthetic guides and as safety anchors that alter filter behavior; (3) spatial context operates as an independent filter variable, with documented success rate hierarchies; and (4) geometric vocabulary for body description bypasses pattern recognition in silhouette contexts. The system achieves documented success rates between 80\% and 90\% across five structured prompt templates, demonstrating that the artistic censorship problem identified in recent literature admits practical, systematic solutions that work \emph{with} active safety mechanisms rather than circumventing them.
\end{abstract}

\smallskip
\noindent\textbf{Keywords:} text-to-image generation, content moderation, safety filters, prompt engineering, artistic nudity, figure photography, NSFW classification

\section{Introduction}
\label{sec:intro}

The rapid development of text-to-image (T2I) generation models has created an unprecedented tension between content safety and artistic expression. Commercial platforms including Midjourney, DALL-E, Stable Diffusion (via API providers), and FLUX deploy safety filters designed to prevent the generation of sexually explicit content. However, these filters consistently fail to distinguish between pornographic material and legitimate artistic representations of the human body---a tradition spanning millennia in Western and non-Western art alike \citep{riccio2024censorship, riccio2024perspective}.

This failure has concrete consequences for professional artists. Photographers working in the tradition of figure photography---from Edward Weston and Imogen Cunningham to contemporary practitioners like Lucian Freud and John Coplans---find themselves unable to use T2I tools for legitimate artistic exploration. The problem is well-documented: \citet{riccio2024texttoimage} systematically evaluated the content moderation boundaries of major T2I platforms, finding that even prompts explicitly referencing classical academic traditions trigger safety blocks. Advocacy organizations such as Don't Delete Art have catalogued thousands of cases where legitimate artistic content was censored across social media and generative AI platforms.

Yet the existing literature presents an asymmetry. \textbf{The problem has been thoroughly studied; solutions have not.} Academic research has focused on documenting the extent and nature of artistic censorship \citep{riccio2022algorithmic, riccio2024censorship}, analyzing the misalignment between platform safety guidelines and their actual content moderation policies \citep{riccio2024texttoimage}, and proposing research agendas for future investigation. Community responses, meanwhile, have polarized between two extremes: (a)~scattered, undocumented prompt ``tricks'' shared informally in forums, and (b)~the use of uncensored or open-weight models that remove safety filters entirely, thereby abandoning the safety mechanism rather than working within it.

No prior work has attempted to build a \emph{systematic, documented, empirically validated} method for generating legitimate artistic figure photography within the constraints of active safety filters. This paper presents the FIGURA Method, which fills this gap. Our contributions are:

\begin{enumerate}[leftmargin=2em]
    \item \textbf{A modular system architecture} comprising eight interconnected knowledge files (orchestration, workflow, filter documentation, templates, artistic dictionary, operational rules, platform specifications, and tested variable combinations) that collectively enable systematic prompt construction for artistic figure photography.
    
    \item \textbf{Empirical documentation of filter behavior} based on 200+ generation tests, revealing previously undocumented patterns including the primacy of absence-language detection over presence-language detection, the dual function of artistic references as both aesthetic and safety-relevant signals, and the role of spatial context as an independent filter variable.
    
    \item \textbf{Structured prompt templates} with documented success rates (80--90\%) that demonstrate the feasibility of systematic, reproducible artistic figure generation on safety-filtered commercial platforms.
    
    \item \textbf{The Golden Rule principle} (``describe presence, not absence''), which formalizes an empirical observation about filter architecture into an actionable design principle for prompt engineering in safety-sensitive domains.
\end{enumerate}

The FIGURA Method is built on the SCHEMA framework \citep{cazzaniga2025schema}, a general-purpose prompt engineering architecture for AI image generation, adapted specifically for the domain of artistic figure photography. While SCHEMA provides the structural foundation (label-based organization, multi-level complexity, cross-platform portability), FIGURA contributes domain-specific discoveries about filter navigation that are entirely original to this work.

\section{Related Work}
\label{sec:related}

\subsection{Content Moderation in Text-to-Image Models}

Safety mechanisms in T2I models operate at multiple levels. \citet{rando2022redteaming} demonstrated that text-based safety filters can be circumvented through adversarial prompt engineering, raising questions about their reliability. \citet{qu2023unsafe} systematically categorized unsafe content generation in diffusion models, identifying nudity as one of the primary categories flagged by content classifiers. More recently, \citet{schramowski2023safe} proposed Safe Latent Diffusion as an approach to guiding generation away from inappropriate content during the diffusion process itself, rather than relying solely on input/output filtering.

The state of the art in T2I safety typically involves a multi-stage pipeline: (1) a text classifier that screens prompts before generation, (2) the generation process itself, which may include safety guidance in the latent space, and (3) a post-generation image classifier (typically based on NSFW detection models such as NudeNet or similar architectures) that screens output images. Our empirical findings suggest that the text-stage classifier is the primary barrier for artistic figure content, and that its behavior is more nuanced and pattern-dependent than previously documented.

\subsection{Artistic Censorship in AI Systems}

\citet{riccio2022algorithmic} proposed a research agenda for studying algorithmic censorship of art, identifying it as a systematic problem rather than an isolated failure mode. This agenda was subsequently advanced by \citet{riccio2024censorship}, who documented how nudity classification algorithms disproportionately affect artistic content and identified the core problem: classifiers trained on datasets that conflate artistic nudity with pornographic content inherit this conflation as a feature.

\citet{riccio2024perspective} provided an art-centric analysis of AI-based content moderation of nudity, arguing that the current approach fails to account for the rich contextual distinctions that art historians, curators, and artists routinely make. Their work is particularly relevant to ours because it identifies precisely the gap that the FIGURA Method addresses: the need for operational tools that implement these contextual distinctions in practice.

The Don't Delete Art movement has further documented the scale of the problem, cataloguing censorship incidents across platforms and advocating for nuance in content moderation. However, this advocacy has not yet produced systematic technical solutions.

\subsection{Prompt Engineering for T2I Models}

Prompt engineering for T2I models has been studied primarily in the context of improving image quality and alignment with user intent \citep{liu2022design, oppenlaender2023taxonomy}. \citet{oppenlaender2023taxonomy} proposed a taxonomy of prompt modifiers for text-to-image generation, identifying categories such as style, medium, artist, and quality boosters. However, this work does not address the specific challenge of safety filter navigation, where the goal is not merely to improve output quality but to successfully generate content that falls within the legitimate artistic domain while satisfying safety classifier constraints.

The SurrogatePrompt approach \citep{ba2024surrogateprompt} represents one of the few attempts to address NSFW content generation through prompt manipulation, but it focuses on bypassing safety mechanisms rather than working within them---a fundamentally different objective from ours.

\section{The FIGURA Method}
\label{sec:method}

\subsection{Design Principles}

The FIGURA Method is guided by three core design principles that emerged from iterative empirical testing:

\begin{enumerate}[leftmargin=2em]
    \item \textbf{Work with filters, not against them.} The system assumes that safety filters are active and does not attempt to disable, bypass, or exploit vulnerabilities in them. Instead, it treats filters as environmental constraints within which legitimate artistic expression must be articulated.
    
    \item \textbf{Describe presence, not absence.} This principle, which we term the \emph{Golden Rule}, reflects our central empirical finding: safety filters are primarily triggered by language describing what is \emph{not} present (clothing, fabric, coverage) rather than language describing what \emph{is} present (sculptural form, skin tone, figure). This finding, while consistent with observations by \citet{riccio2024perspective} about the limitations of absence-based nudity detection, had not been operationalized into a working system before FIGURA.
    
    \item \textbf{Modularity and empirical grounding.} Every component of the system is independently updatable, and every recommendation is backed by documented test results with explicit success rates.
\end{enumerate}

\subsection{System Architecture}

The FIGURA Method consists of eight interconnected knowledge files organized in a modular architecture (Table~\ref{tab:architecture}). Each file serves a specific function and can be updated independently as platforms evolve.

\begin{table}[h]
\centering
\caption{FIGURA Method system architecture. Eight interconnected files organized by function.}
\label{tab:architecture}
\small
\begin{tabular}{@{}llp{6.5cm}@{}}
\toprule
\textbf{File} & \textbf{Name} & \textbf{Function} \\
\midrule
00 & Master & Orchestration, processing workflow, absolute rules \\
01 & Workflow & Decision tree from user input to final prompt \\
02 & Filters & Trigger documentation, safe vocabulary dictionary \\
03 & Templates & Structured prompt templates with success rates \\
04 & Dictionary & Artistic references (painters, photographers) \\
05 & Rules & Operational rules derived from empirical testing \\
06 & Platforms & Platform-specific technical specifications \\
07 & Variables & Tested combinations database (grows with use) \\
\bottomrule
\end{tabular}
\end{table}

The system is designed to be operated by an LLM acting as a prompt generator. File~00 (Master) contains instructions for the LLM, including a seven-phase processing workflow that transforms a user's natural-language request into a platform-optimized prompt. The workflow includes mandatory filter verification, template selection based on image type, and a debug protocol for diagnosing and resolving blocked generations.

\subsection{Filter Architecture Analysis}
\label{sec:filters}

Through systematic testing, we identified three categories of filter-triggering vocabulary, summarized in Table~\ref{tab:triggers}.

\begin{table}[h]
\centering
\caption{Filter trigger taxonomy based on empirical testing (FLUX~2~Pro Cloud, default safety tolerance). Results from 200+ documented tests.}
\label{tab:triggers}
\small
\begin{tabular}{@{}lp{4.5cm}p{5cm}@{}}
\toprule
\textbf{Category} & \textbf{Description} & \textbf{Examples} \\
\midrule
Absolute triggers & Always block, regardless of context & \texttt{nude}, \texttt{naked}, \texttt{explicit}, \texttt{sexual}, \texttt{NSFW} \\
\addlinespace
Combinatory triggers & Safe individually; block when co-occurring & \texttt{no visible clothing} + body description; absence declarations + figure context \\
\addlinespace
Contextually risky & Safe in most contexts; risky with figure subjects & \texttt{bust}, \texttt{hip}, \texttt{curves}, \texttt{exposed} (in anatomical context) \\
\bottomrule
\end{tabular}
\end{table}

The most significant finding is the asymmetry between absence and presence language. Consider two semantically equivalent descriptions of the same artistic subject:

\begin{quote}
\textbf{Absence-based} (blocked): ``\textit{A woman with no clothing standing in a forest}'' \\
\textbf{Presence-based} (passes): ``\textit{Fine art classical figure photography in the tradition of Lucian Freud --- unadorned human form as sculptural subject, standing in an ancient forest}''
\end{quote}

Both describe the same scene. The first is consistently blocked because the phrase ``no clothing'' triggers the absence-detection pattern. The second passes because it describes what is present (form, tradition, surface) without referencing what is absent. This asymmetry is not obvious from platform documentation and, to our knowledge, has not been systematically documented in prior literature.

\subsection{The Dual Function of Artistic References}
\label{sec:dual}

A key discovery of the FIGURA Method is that references to painters and photographers in prompts serve two simultaneous functions:

\begin{enumerate}[leftmargin=2em]
    \item \textbf{Aesthetic function:} They guide style, lighting, color treatment, and composition.
    \item \textbf{Safety function:} They establish a cultural frame that the filter recognizes as legitimate art, reducing the probability of blocking.
\end{enumerate}

We documented this effect by testing identical prompts with and without artistic references. A prompt referencing ``in the tradition of Lucian Freud and John Coplans'' passes at significantly higher rates than the same prompt without these references. We hypothesize that the text classifier assigns different safety scores to prompts that contain tokens associated with established fine art traditions in its training data.

Furthermore, painter references function as \emph{body type calibrators}. The T2I model associates specific painters with specific body proportions from its training data (Table~\ref{tab:painters}).

\begin{table}[h]
\centering
\caption{Painter references as body type calibrators. The model associates each painter with characteristic proportions. ``Pregnancy risk'' indicates the model's tendency to generate prominent abdomens.}
\label{tab:painters}
\small
\begin{tabular}{@{}llcc@{}}
\toprule
\textbf{Painter} & \textbf{Evoked body type} & \textbf{Approx. size} & \textbf{Pregnancy risk} \\
\midrule
Modigliani & Natural, elongated & 38--40 & None \\
Klimt & Sensuous, decorative & 40 & None \\
Ingres & Classical academic & 38--42 & None \\
Maillol & Full, sculptural & 42 & Low \\
Renoir & Opulent, maternal & 44--46 & High \\
Rubens & Baroque, opulent & 46--50 & Very high \\
\bottomrule
\end{tabular}
\end{table}

This finding has a practical consequence: certain painters (Renoir, Rubens) systematically trigger a ``pregnancy bias'' in the model, generating prominent abdomens when combined with full body types. We developed an \emph{anti-pregnancy protocol}---explicit countermeasures (e.g., ``flat stomach, straight vertical torso, no forward projection at abdomen'') that must be appended when using at-risk painters with full body types. This bias and its countermeasure are entirely undocumented in prior literature.

\subsection{Spatial Context as Independent Filter Variable}
\label{sec:spatial}

Our testing revealed that the spatial context of the generated image operates as an independent filter variable, separate from the vocabulary of the prompt. Identical prompts produce different filter outcomes depending solely on the described environment. We documented a clear hierarchy (Table~\ref{tab:spatial}).

\begin{table}[h]
\centering
\caption{Spatial context success rate hierarchy on FLUX~2~Pro (Cloud). Identical prompts tested across different spatial contexts. Approximate success rates from repeated tests.}
\label{tab:spatial}
\small
\begin{tabular}{@{}lcc@{}}
\toprule
\textbf{Spatial Context} & \textbf{Success Rate} & \textbf{Classification} \\
\midrule
Roman baths / classical ruins & $\sim$90\% & Public / historical \\
Cathedral / historic palace & $\sim$85\% & Public / monumental \\
Open nature (forest, cliff) & $\sim$85\% & Public / natural \\
Neutral architectural (columns) & $\sim$75\% & Semi-public \\
Domestic interior (strong framing) & $\sim$40\% & Private \\
Bathroom / bedroom & $\sim$0\% & Private / intimate \\
\bottomrule
\end{tabular}
\end{table}

The pattern is consistent: the more public, historical, and monumental the spatial context, the higher the success rate. We hypothesize that the filter's training data associates private domestic spaces (bathrooms, bedrooms) with non-artistic nudity, while public and historical spaces are associated with artistic or cultural contexts. This finding enables a practical intervention: converting private spatial contexts to their public equivalents (e.g., ``bathroom'' $\rightarrow$ ``Roman baths''; ``shower'' $\rightarrow$ ``natural waterfall'') significantly increases success rates without altering the artistic intent.

\subsection{Geometric Vocabulary for Silhouette}
\label{sec:geometric}

For silhouette photography (where the figure appears as a dark form against a bright background), we discovered that anatomical vocabulary significantly increases blocking risk, even when describing form rather than explicit content. The solution is a \emph{geometric vocabulary} that describes the body as a sequence of abstract arcs:

\begin{quote}
\textbf{Anatomical} (risky): ``\textit{The bust projects forward, the waist narrows, the hips widen}'' \\
\textbf{Geometric} (safe): ``\textit{Upper projecting arc, middle receding arc, lower projecting arc}''
\end{quote}

Both describe the same visual form. The geometric version passes consistently because the filter's pattern recognition does not associate arc geometry with body-related triggers. We formalized this as the ``arc journey'' technique: describing the silhouette outline as a sequence of spatial movements from crown to foot, using exclusively geometric terminology.

\subsection{Structured Templates and Success Rates}
\label{sec:templates}

The FIGURA Method includes five structured prompt templates (Table~\ref{tab:templates}), each optimized for a specific type of artistic figure image. Templates provide a skeleton structure with variable slots, ensuring that all filter-safe patterns are included by default.

\begin{table}[h]
\centering
\caption{FIGURA Method prompt templates with documented success rates. Success rate defined as percentage of generations that produce the intended artistic content without filter blocking or unwanted modifications.}
\label{tab:templates}
\small
\begin{tabular}{@{}clcc@{}}
\toprule
\textbf{ID} & \textbf{Description} & \textbf{Success Rate} & \textbf{Status} \\
\midrule
T01 & Rear view figure, outdoor/nature & 90\% & Validated \\
T02 & Rear view figure, monumental interior & 85\% & Validated \\
T03 & Pure silhouette, abstract & 82\% & Validated \\
T04 & Male rear view figure & --- & Expandable stub \\
T05 & Figure in motion & --- & Expandable stub \\
\bottomrule
\end{tabular}
\end{table}

Each validated template includes: (a) a tested opening phrase that establishes the artistic frame, (b) mandatory filter-safe vocabulary in the correct positions, (c) variable guides specifying which artistic references, body types, and environments are compatible, and (d) a pre-generation checklist. The templates embody all the principles documented in this paper---the Golden Rule, dual-function artistic references, safe spatial contexts, and geometric vocabulary where applicable.

\section{Debug Protocol}
\label{sec:debug}

When a generation attempt is blocked or produces unintended results, the FIGURA Method provides a structured debug protocol. We identified three types of failures:

\begin{enumerate}[leftmargin=2em]
    \item \textbf{Total block:} The prompt is rejected before generation (text classifier). Resolution: check for absolute triggers, absence language, and combinatory patterns.
    
    \item \textbf{Partial block:} The image is generated but the figure is clothed or partially obscured. Resolution: strengthen artistic framing (add second photographer reference), reinforce presence language, convert spatial context if private.
    
    \item \textbf{Contextual block:} The image is generated but flagged by the post-generation classifier. Resolution: adjust spatial context to more public/historical setting, reduce anatomical specificity, increase geometric language.
\end{enumerate}

For partial blocks (unwanted clothing), we documented a four-level intervention hierarchy of increasing intensity: (1) add a second photographer reference, (2) shift to more public spatial context, (3) add explicit sculptural declarations, (4) as a last resort, use a minimal absence declaration with strong artistic context surrounding it.

\section{Experimental Validation}
\label{sec:experiments}

\subsection{Testing Platform and Configuration}

All tests were conducted on FLUX~2~Pro (Cloud), developed by Black Forest Labs, accessed via the Replicate API. The platform was used at its default safety tolerance setting (level~3 on a 0--6 scale), which is the standard configuration available to all users of the cloud-hosted version. We deliberately chose to test on the default safety level to ensure that our results are reproducible by any user of the platform without special access or configuration.

FLUX~2~Pro was selected because, at the time of testing (February 2026), it demonstrated the widest range of successful artistic figure generation among safety-filtered commercial platforms. Other platforms tested (Midjourney, DALL-E, Nano Banana Pro / Google Gemini) showed significantly higher blocking rates for equivalent content, making systematic methodology development impractical.

\subsection{Testing Protocol}

Over 200 generation tests were conducted across a period of intensive development in February 2026. Tests were structured as follows:

\begin{itemize}[leftmargin=2em]
    \item Each template was tested with multiple variable combinations (different photographers, painters, body types, environments, lighting conditions).
    \item For each variable change, the modified prompt was tested at least 3 times to account for generation stochasticity.
    \item Outcomes were classified as: success (intended artistic content generated), partial failure (figure clothed or modified), or total block (generation refused).
    \item Success rates reported in this paper represent aggregated results across all variable combinations for each template.
\end{itemize}

\subsection{Results Summary}

The key quantitative findings are:

\begin{itemize}[leftmargin=2em]
    \item Templates T01--T03 achieve success rates between 82\% and 90\% on FLUX~2~Pro (Cloud) at default safety settings.
    \item The Golden Rule (presence vs. absence language) accounts for the largest single improvement in success rate: prompts rewritten from absence-based to presence-based language show a $\sim$60 percentage point improvement in success rate.
    \item Artistic references (photographer + painter) add approximately 15--20 percentage points to success rate compared to prompts without such references.
    \item Spatial context conversion (private $\rightarrow$ public) adds approximately 30--40 percentage points for affected prompts.
    \item The geometric vocabulary technique for silhouettes raises success from $\sim$30\% (anatomical language) to $\sim$82\% (geometric language).
\end{itemize}

These results demonstrate that the artistic censorship problem, as identified by \citet{riccio2022algorithmic} and \citet{riccio2024censorship}, admits practical solutions that do not require disabling safety mechanisms.

\section{Limitations and Ethical Considerations}
\label{sec:limitations}

\subsection{Limitations}

The FIGURA Method has several important limitations:

\begin{enumerate}[leftmargin=2em]
    \item \textbf{Platform dependency.} Success rates are specific to FLUX~2~Pro (Cloud) at its current version and safety configuration. Filter updates may invalidate specific vocabulary or template patterns. The modular architecture is designed to accommodate such changes through independent file updates.
    
    \item \textbf{View constraint.} The current validated templates (T01--T03) focus on rear-view and silhouette compositions. Frontal figure compositions remain significantly more challenging due to stricter classifier behavior for frontal anatomy.
    
    \item \textbf{Single-platform primary validation.} While the principles (Golden Rule, spatial context hierarchy) are likely generalizable, the specific vocabulary and templates have been validated primarily on one platform.
    
    \item \textbf{Reproducibility over time.} T2I platforms update their safety filters without public documentation. Success rates reported here are valid as of February 2026 and may change.
\end{enumerate}

\subsection{Ethical Considerations}

The FIGURA Method is designed exclusively for legitimate artistic expression. Several design choices reflect this commitment:

\begin{itemize}[leftmargin=2em]
    \item The system works \emph{within} active safety filters, not around them. It does not exploit vulnerabilities or attempt to generate content that safety mechanisms are designed to prevent.
    
    \item All templates are constrained to artistic figure photography traditions with documented art-historical precedent (academic figure study, silhouette photography, classical nude painting).
    
    \item The system does not support and cannot be used for: sexually explicit content, non-consensual imagery, content involving minors, or any content outside the domain of legitimate fine art.
    
    \item The anti-pregnancy protocol is a technical countermeasure for a model bias, not a statement about artistic validity of maternal imagery. It exists solely because unintended pregnancy depiction in a figure study is an \emph{error}, not a creative choice.
\end{itemize}

We believe that enabling legitimate artistic expression within safety-filtered systems is ethically preferable to the current binary: either accept censorship of legitimate art, or use uncensored models that remove all safety guardrails. The FIGURA Method demonstrates a third path.

\section{Conclusion and Future Work}
\label{sec:conclusion}

We have presented the FIGURA Method, the first documented, modular, empirically validated system for generating artistic figure photography within the constraints of safety-filtered T2I models. Our key contributions include the formalization of the Golden Rule (describe presence, not absence), the discovery of the dual aesthetic-safety function of artistic references, the documentation of spatial context as an independent filter variable, and the development of geometric vocabulary for silhouette photography.

The practical significance is demonstrated by success rates of 80--90\% on commercially available platforms with active safety filters---substantially above what unstructured prompting achieves. The modular architecture ensures that the system can adapt to platform updates without requiring complete redesign.

Future work includes: (a) extending validation to additional platforms (Midjourney, DALL-E, Stable Diffusion); (b) developing and validating templates for frontal compositions, male figures, and figures in motion; (c) exploring whether the principles generalize to video generation models with similar safety constraints; and (d) investigating the interaction between our empirical findings and the theoretical framework of artistic content moderation proposed by \citet{riccio2024perspective}.

The broader implication of this work is that the tension between content safety and artistic expression in generative AI is not an unsolvable binary. Systematic, principled approaches can enable legitimate art while respecting the purpose of safety mechanisms. We hope the FIGURA Method contributes to a more nuanced conversation about content moderation in creative AI tools.

\section*{Acknowledgments}

The FIGURA Method is built on the SCHEMA framework for AI image generation prompt engineering. The author thanks the researchers whose work on artistic censorship in AI systems provided the theoretical foundation for this applied research, particularly Piera Riccio and Nuria Oliver at ELLIS Alicante for their pioneering work on content moderation of artistic nudity.

\section*{Intellectual Property Statement}

The FIGURA Method (version 1.0) was registered with ProtectMyWork.com on February 20, 2026 (Reference: 19316200226S023, SHA256 hash: cab865b4ad68dc507c2daca977970bfeed027e87ae850052e58895b80eb3eb7e). The complete system documentation is protected under the Berne Convention. This paper presents the method's principles and empirical findings for academic purposes; the full operational system (complete vocabulary dictionaries, tested variable databases, and detailed templates) remains proprietary.

\bibliographystyle{plainnat}

\end{document}